\documentclass{article}

\PassOptionsToPackage{numbers}{natbib}

\usepackage{url}

\usepackage[preprint]{setup}

\usepackage[utf8]{inputenc} 
\usepackage[T1]{fontenc}    
\usepackage{hyperref}       
\usepackage{url}            
\usepackage{booktabs}       
\usepackage{amsfonts}       
\usepackage{nicefrac}       
\usepackage{microtype}      
\usepackage{xcolor}         
\usepackage{float}
\usepackage{tabularx}
\usepackage{rotating}
\usepackage{placeins}
\usepackage{csquotes}
\title{Unbounded Harms, Bounded Law: \\Liability in the Age of Borderless AI}

\author{%
  Ha-Chi Tran\\
  The London School of Economics and Political Science\\
  London, United Kingdom\\
  \texttt{c.h.tran@lse.ac.uk} \\
}

\begin{document}

\maketitle

\begin{abstract}
The rapid acceleration of artificial intelligence (AI) has exposed fundamental deficiencies in prevailing risk governance: despite substantial advances in ex ante harm identification and prevention, Responsible AI scholarship remains comparatively underdeveloped in its treatment of ex post risk governance. Core legal questions regarding compensation, mitigation, attribution of responsibility, liability allocation, and the effectiveness of remedial mechanisms remain inadequately theorized and underinstitutionalized, particularly in relation to transboundary AI harms, AI-induced risks, and damages that transcend national borders, legal jurisdictions, and regulatory authority. Drawing on contemporary AI risk analyses, we argue that such harms are not exceptional but structurally inherent to the AI supply chains, and are likely to increase in both frequency and severity due to globalized deployment, cross-border data infrastructures, and asymmetric national capacities in AI development and oversight, thereby rendering territorially bounded liability regimes increasingly inadequate. Adopting a comparative and interdisciplinary approach, the paper examines compensation and liability mechanisms from adjacent high-risk and transnational domains, including vaccine injury compensation, financial systemic risk governance, commercial nuclear energy liability, and international environmental harm regimes, to identify transferable legal design principles, such as strict liability, pooled compensation mechanisms, and collective risk-sharing, as well as the structural constraints limiting their application to AI-related harms. Situated within an international order increasingly shaped by AI arms race dynamics rather than cooperative governance, the paper outlines the contours of a global AI compensation and accountability architecture, underscoring the tension between geopolitical rivalry and the collective action required to govern transboundary AI risk.
\end{abstract}

\section{Introduction}
The black-boxed, diffuse, and distributed nature of contemporary artificial intelligence (AI) systems, particularly those based on sub-symbolic architectures such as deep neural networks, poses fundamental challenges for legal systems, especially with respect to risk governance and liability allocation \citep{r:84, r:63, r:82, r:83}. To date, the rapidly expanding discourse on Responsible AI has been dominated by ex ante approaches to risk management. Scholarly and regulatory efforts have prioritized preventive tools such as risk taxonomy and impact assessments \citep{r:3, r:4, r:5, r:6}, technical and organizational safeguards \citep{r:7, r:8}, and algorithmic audits \citep{r:9, r:10} designed to promote safety, fairness, and compliance across the AI lifecycle. The European Union’s risk-based regulatory framework under the AI Act \citep{r:23}, and its growing potential influence on AI governance in other jurisdictions through the Brussels Effect \citep{r:24, r:25}, exemplifies this preventive paradigm by classifying AI systems according to perceived risk and subjecting them to differentiated regulatory obligations.

However, the intrinsic characteristics of contemporary AI systems—most notably their emergent behaviors, adaptive learning capacities, reliance on continuously evolving datasets, and context-dependent performance—fundamentally limit the effectiveness of predictive, ex ante risk governance \citep{r:17, r:85, r:86}. A growing body of legal scholarship has accordingly exposed the structural inadequacy of prevailing liability regimes, including both strict and fault-based frameworks, in addressing AI-related harms \citep{r:11, r:12, r:13, r:14, r:15}. The defining features of modern AI systems—such as epistemic opacity, autonomous and adaptive decision-making, distributed and modular functionality, and complex multi-actor development and deployment chains—destabilize the core doctrinal assumptions underpinning traditional liability law. In particular, they challenge foundational premises of linear causation, demonstrable negligence, and the clear attribution of product defects in design, warning, or operation \citep{r:16, r:17, r:12, r:80}.

These structural mismatches generate persistent evidentiary asymmetries that systematically disadvantage claimants, thereby constraining access to compensation and undermining the corrective, allocative, and deterrent functions that liability law traditionally serves. Legal frameworks premised on the assumption of relatively stable technologies, traceable causal chains, and identifiable human agency are thus increasingly ill-equipped to address harms arising from AI systems. Even highly sophisticated ex ante risk taxonomies and compliance mechanisms cannot reliably anticipate emergent or downstream harms that materialize only through real-world deployment and interaction \citep{r:12, r:52}. This irreducible uncertainty, inherent in AI, a technology that is actively reshaping global socio-economic trajectories, has in turn catalysed a shift toward adaptive governance instruments—most notably regulatory sandboxes \citep{r:26, r:87}—which enable regulators and firms to engage in iterative learning through controlled experimentation and empirical observation, thereby surfacing risks that evade conventional predictive assessment and audit \citep{r:26, r:27}.

At a systemic level, the emergence and increasing reliance on adaptive governance tools reflect a deeper condition of epistemic uncertainty in global AI governance. Not only is knowledge about the nature of potential AI risks inherently incomplete, but the probability and severity of those risks are likewise indeterminate, a Knightian uncertainty \citep{r:28}. In such a governance environment, an exclusive emphasis on ex ante risk prevention is insufficient. Instead, greater attention must be directed toward ex post governance mechanisms capable of responding once harm has occurred. Despite this necessity, legal and policy scholarship has devoted comparatively little attention to the ex post dimensions of AI governance \citep{r:88}. Key post-harm questions remain underexamined: who is responsible for AI-induced harm, what constitutes fair and effective compensation, and whether current remedies can deliver timely redress while mitigating future risks.

Recent legal reforms, most notably within the European Union, including the revised Product Liability Directive and the proposed (but ultimately unadopted) AI Liability Directive, reflect a growing institutional recognition of a structural liability gap in the governance of AI-related harms, particularly by explicitly acknowledging the evidentiary and procedural disadvantages faced by individuals harmed by AI systems \citep{r:29, r:30}. When read alongside an expanding body of broader scholarly proposals advocating several liability architectures modifications, these developments point to an emerging consensus that conventional liability paradigms are increasingly ill-suited to addressing harms generated by autonomous and adaptive AI systems. By recalibrating the role of fault, lowering evidentiary thresholds, and reallocating elements of the burden of proof to developers and deployers—most notably through the introduction of rebuttable presumptions or per se rule of AI defectiveness and dangerousness to shift the burden of proof away from the claimants \citep{r:1}—these reforms and proposals collectively signal an incipient doctrinal shift toward stricter for high-risk systems \citep{r:31}, and in certain contexts, no-fault liability models \citep{r:11, r:12, r:13}. This shift is normatively oriented toward restoring the compensatory, corrective, and risk-allocative functions of liability law under conditions of technological opacity, autonomy, and adaptive learning.

Notwithstanding these important legal reforms, prevailing debates on AI liability remain confined mainly within nationally bounded regimes. This orientation is consistent with the territorial logic of legal sovereignty that underpins the contemporary international legal order and reflects states’ enduring preference to retain regulatory authority over emerging technologies within their own jurisdictions. Yet such a territorially constrained focus risks obscuring a critical and increasingly salient category of AI-related risk: transboundary, regional, and global harms. These encompass injuries and losses arising from AI systems that are developed, trained, deployed, and operated across multiple jurisdictions, as well as harms whose scale, diffusion, or systemic character exceeds the regulatory, adjudicative, and compensatory capacities of any single state. This paper seeks to address this gap by advancing a conceptual and comparative analysis of transboundary AI risk and the governance challenges it poses for ex post compensation and insurance. Specifically, the analysis focuses on third-party harms suffered by affected individuals and communities, distinct from losses borne by AI developers or system owners and deployers, thereby foregrounding the limits of territorially bounded liability regimes in an increasingly transnational AI ecosystem.

The paper proceeds as follows. Section II develops a conceptual framework for transboundary AI risk, highlighting how AI’s pervasive multi-use across sectors and its potential for large-scale, long-term harm generate cross-border risks that differ in complexity, temporal scope, and severity. It argues that globalized deployment, cross-border data infrastructures, and asymmetries in national regulatory capacity are likely to intensify both the frequency and magnitude of such risks, rendering territorially bounded liability and compensation regimes increasingly inadequate and calling for differentiated, layered governance responses. Section III offers a comparative analysis of compensation and risk-sharing mechanisms in other transnational domains, with particular attention to hybrid public–private arrangements for insurance and reinsurance. Drawing on cases ranging from COVID-19 vaccine injury compensation to public–private schemes in nuclear energy, financial systemic risk, and environmental governance, it examines how institutional designs allocate responsibility, pool risk, and manage uncertainty and collective action problems. These comparisons underscore both the potential and the limits of such mechanisms, especially where geopolitical rivalry and divergent economic incentives constrain durable international cooperation.

Based on this comparative analysis, the paper distills a set of legal design principles potentially relevant for cross-border AI governance, including strict liability, collective compensation funds, liability channeling, and pooled risk-sharing arrangements, while critically examining the structural, political, and institutional constraints that complicate their implementation in an international order increasingly shaped by AI arms race dynamics rather than cooperative governance \citep{r:89, r:90}. The analysis outlines the contours of a potential global architecture for third-party AI harm compensation and accountability, highlighting the persistent tension between geopolitical rivalry and the collective action required to manage transboundary AI risks. Yet, the paper is deliberately conceptual and diagnostic rather than prescriptive: given the substantial political and economic barriers to international consensus on binding liability regimes, it does not propose a specific institutional design but instead seeks to clarify the problem space, map relevant governance analogies, and identify conditions under which transnational AI compensation mechanisms might plausibly emerge.

\section{Epistemic Uncertainty: The Complex Transnational Dynamics of AI Harms}

One of the foundational conditions of effective risk governance is the capacity to classify risks, pool risks, and render uncertainty at least partially intelligible through probabilistic estimation of likelihood, modes of occurrence, and potential harm \citep{r:2, r:32, r:33, r:34}. Historically, particularly within insurance and reinsurance regimes, this capacity has been central to the transformation of ex post uncertainty into ex ante anticipation and preparedness \citep{r:33, r:2, r:37}. By translating uncertain and dispersed harms into calculable and collectively borne risks, insurance-based mechanisms have enabled proactive responses to harmful events and institutionalized predictable pathways for compensation and redress.

Yet, the extent to which such mechanisms can successfully operate varies markedly across risk domains, largely as a function of complexity, temporal horizon, and systemic interdependence. In areas such as individual healthcare, public and private insurance arrangements have matured to provide relatively comprehensive coverage and routinized compensation \citep{r:2}. By contrast, risks that are foreseeable but collective, systemic, and temporally diffuse, such as those associated with climate change or large-scale environmental degradation, continue to exceed the absorptive capacity of existing insurance and reinsurance markets \citep{r:2}. In these contexts, the limits of market-based risk governance have necessitated cross-sectoral and transnational interventions, which frequently encounter entrenched political–economic resistance and coordination failures.

This section, thus, develops a conceptual taxonomy to systematize how cross-border AI-related harms are likely to manifest in practice. Drawing on empirical cases as well as forward-looking analyses informed by contemporary AI risk-mapping efforts, it elucidates the forms, transmission pathways, and temporal dynamics of such harms. The core claim advanced is that, unlike more mature sectors such as finance, healthcare, or personal property damage, where risks tend to cluster within a relatively narrow set of categories defined by recognizable complexity and time horizons, AI’s inherently multi-use and general-purpose character enables it to generate risks across a far more heterogeneous and non-static shifting spectrum.

\subsection{Complexity and Differentiation in Transnational AI Risk}

Cross-border AI harms and risks are those whose impacts, scale, or causal chains extend beyond the territorial boundaries of a single state, implicating actors, individuals, and interests across multiple jurisdictions. Such harms manifest across varying magnitudes. At the micro level, they may involve individualized injuries with transnational legal implications. An example is an Australian regional mayor who contemplated legal action against the developer of the AI chatbot ChatGPT, alleging defamation after the system, developed by OpenAI, a U.S.-based company, falsely implicated him in a foreign bribery scandal \citep{r:18}. Although the claim did not ultimately proceed, it exposed a dense set of legal challenges, including the attribution of harm to AI-mediated intermediaries and the determination of jurisdiction and applicable law, thereby exemplifying the structural complexity of cross-border AI disputes even where harms are narrowly individualized.

At a more systemic level, cross-border harms from AI arise from the transnational deployment of AI systems embedded in globally fragmented value chains. For example, a firm domiciled in a Southeast Asian jurisdiction may rely on open-source foundational models developed by U.S.-based technology companies (E.g., some Llama versions of Meta) to build and deploy a medical advisory chatbot. If the system disseminates inaccurate health information, liability may be relatively clearer with respect to the deploying entity within the state's territory, yet the potential responsibility of foreign upstream providers of foundational models remains legally unsettled, highlighting persistent difficulties in allocating responsibility across multi-actor AI supply chains that span multiple legal orders.

The risks of cross-border AI also include potentially catastrophic harms. Although their likelihood, causal pathways, and temporal dynamics remain contested within both technical and policy communities, their defining characteristic is that, if realized, their consequences would exceed the regulatory, remedial, and institutional capacities of any single state \citep{r:35, r:36}. Such harms may materialize abruptly, for example, AI-enabled military or nuclear systems autonomously initiating or escalating conflicts, cyber-attacks, or may unfold more gradually through long-term cultural, informational, or environmental externalities associated with the large-scale development, training, and deployment of AI systems. In both cases, the transboundary character of these risks fundamentally challenges territorially bounded models of risk governance, liability, and compensation, rendering purely national responses structurally inadequate.

In addressing AI-related harms, whether territorial or transboundary, scholars and policymakers have increasingly relied on classificatory frameworks to pool risk and render it governable \citep{r:2}. These range from sector- or domain-based typologies to severity-based risk tiering, exemplified by the EU AI Act. A further analytically and operationally productive approach, suggested by Louis W. Pauly, classifies risks according to their complexity and the temporal horizon required for effective response, remediation, and compensation \citep{r:2}. From this perspective, risks can be organized into four stylized categories: (1) low-complexity risks with clearly identifiable victims, locations, and quantifiable harms, and short time horizons for intervention; (2) low-complexity risks whose effects unfold over extended periods, such as nuclear generator accidents; (3) high-complexity risks demanding rapid, coordinated responses, including global financial market crises; and (4) high-complexity risks with long-term, diffuse, and systemic impacts, such as environmental degradation and related collective harms (see Figure 1).

\begin{figure}[h]
  \centering
  \includegraphics[width=0.8\linewidth]{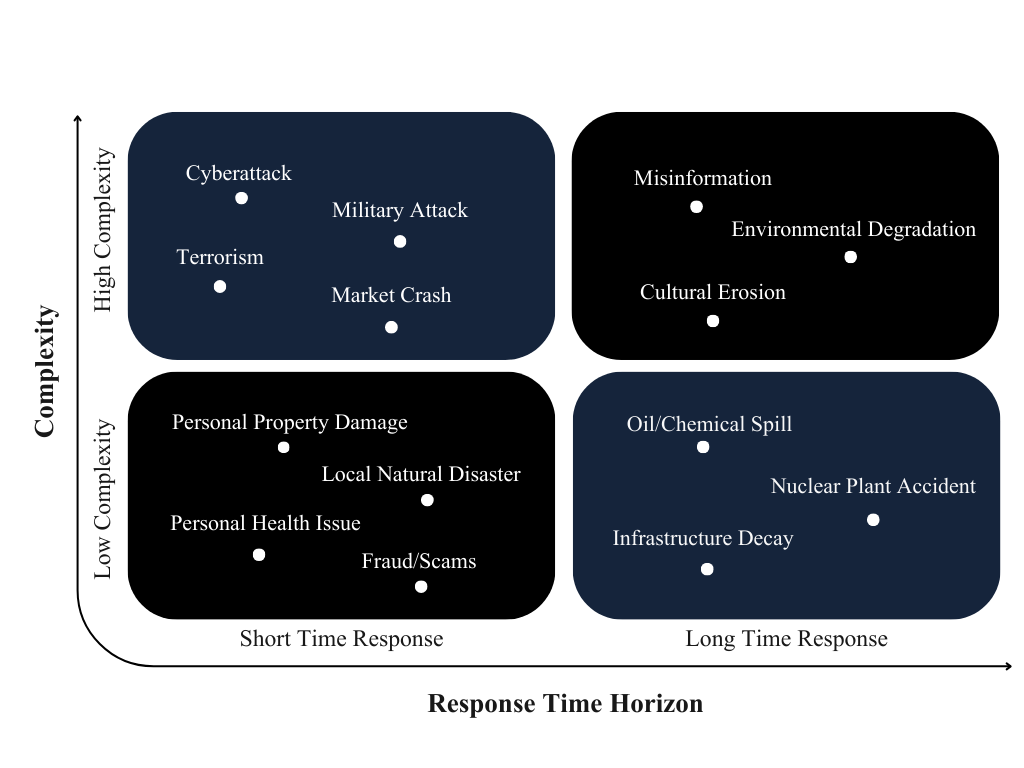}
  \caption{Risk Matrix (Relative): Complexity vs. Response Time}
\end{figure}

Although schematic, this typology illuminates the uneven maturity and preparedness of existing risk-management, insurance, and reinsurance systems, as well as the fragmented and uncertain character of the global risk governance ecosystem, particularly in public–private interactions. Risks in the low-complexity categories have historically been most amenable to private market solutions \citep{r:2}. AI-related harms in this category, while introducing novel uncertainties, are increasingly incorporated into emerging insurance products. Recent debates over insurance coverage for financial losses arising from generative AI, especially chatbot hallucinations, illustrate this trend, with major insurers piloting policies specifically designed to address such risks \citep{r:19, r:20}. The defining feature of this category is the relative ease of identifying affected parties and quantifying harm, which enables actuarial assessment and incentivizes efficient market development. Over time, private insurance mechanisms have matured to price, pool, and compensate for such harms, with the state confined mainly to a regulatory role aimed at safeguarding market integrity and preventing anti-competitive conduct in the context of non-catastrophic losses \citep{r:2, r:38}. For more catastrophic risks, such as nuclear generator accidents, private insurance may still play a role, but state involvement, or sometimes acting as a reinsurer of last resort,  is often required to exceed coverage ceilings \citep{r:2, r:39, r:40}.

From a policy perspective, particularly amid the gradual shift in liability discourse from fault-based to stricter liability regimes, the expansion of private insurance offers AI firms a practical means to compensate third-party victims, and, in some cases, their own losses, without resolving complex fault attribution across distributed AI value chains. While such arrangements do not determine the more profound doctrinal questions of responsibility allocation in multi-actor ecosystems, they provide more accessible and expedient pathways to compensation, reinforcing the corrective and allocative functions of liability law.

Yet when applied in a cross-border context, even these comparatively tractable risk categories encounter profound governance challenges. Jurisdictional uncertainty, the application of lex loci delicti/damni (the law of the place), divergences in compensation standards, and wide disparities in institutional and judicial capacity across states complicate the functioning of private insurance mechanisms. In jurisdictions with underdeveloped or weakly regulated insurance markets, reliance on market-based solutions risks entrenching inequalities in access to compensation and redress. Here, state intervention, or the participation of intergovernmental or international institutions, may be necessary. Analogous mechanisms, such as Vaccine Injury Compensation Programs and Nuclear Liability Conventions, illustrate how public or quasi-public schemes can fill gaps left by private markets. However, extending such frameworks to AI raises unresolved normative and political–economic questions, particularly concerning the justification for deploying public resources to underwrite private technological risk in domains that do not clearly generate public goods, as in vaccination or renewable energy, or where harms may exceed the financial capacity of private actors, as in nuclear-scale disasters.

For high-complexity, rapid-onset risks, such as global financial crises, public debt defaults, flash market collapses, sudden systemic shocks, or attacks on critical infrastructure, the central governance challenge lies not only in technical or predictive uncertainty, but in the deeply politicized nature of both the risks themselves and the institutional responses they provoke. The economic, strategic, and security sensitivities of such events create strong incentives for states to act unilaterally or opportunistically, relying on discretionary and ad hoc measures \citep{r:2} that may conflict with or interact unpredictably across borders. At the same time, the immediacy and catastrophic potential of these risks can generate powerful incentives for urgent coordination, as the costs of inaction or fragmented responses become immediately apparent.

Historical precedents can be observed in post-financial-crisis interventions, during which multilateral institutions such as the International Monetary Fund were empowered to stabilize exchange rates, coordinate reserve policies, and facilitate collective fiscal and monetary responses \citep{r:41}. Similar mechanisms aimed at rapid, coordinated responses can be found in non-financial contexts; for instance, the EU Civil Protection Mechanism obliges member states to activate support quickly, share resources, and coordinate relief efforts in emergencies \citep{r:42}. Analogous dynamics may emerge in the AI domain, where erroneous or malicious AI-enabled actions could inflict cross-border civilian or military harm. This might include miscalculations in AI-assisted nuclear command-and-control systems, automated escalations in cyber-physical infrastructure, or AI-facilitated attacks on critical systems, demanding immediate, coordinated responses. In such scenarios, the extreme systemic stakes may create a narrow but consequential window in which unilateral action is insufficient, thereby motivating cooperative safeguards or legally binding risk-limitation frameworks, akin to IMF mechanisms for stabilizing the international financial system during liquidity crises or to coordinated civil-military emergency protocols.

By contrast, high-complexity, long-horizon risks, gradual, diffuse, and often non-urgent, present fundamentally different governance challenges. These harms, such as environmental degradation, biodiversity loss, or protracted systemic failures, demonstrate persistent misalignment between the spatial and temporal scale of the risk and the incentives available to individual states or private actors \citep{r:2}. AI analogues include long-term ecological impacts of energy-intensive computation, persistent algorithmic biases that influence social and cultural norms, and the gradual erosion of informational and epistemic integrity in global knowledge networks \citep{r:43, r:44}. In such cases, the diffuse nature of harm, the delayed manifestation of damages, and the presence of competing short-term economic or political incentives, such as prioritizing productivity, innovation, or national competitiveness, systematically undermine investment in mitigation or preventive measures. Both state and private actors may rationally underinvest, relying on others to bear risk or assuming that harms are too diffuse or uncertain to warrant immediate intervention. 

\subsection{From Complexity to Crisis: Amplification and Escalation of Transnational AI Risks}
While the fourfold typology of risk complexity and temporal horizon highlights the differentiated, and sometimes conflicting, incentives and mechanisms available to public and private actors, AI presents a uniquely intractable case. Its constitutive features, multi-use adaptability, epistemic opacity, and embedding in globally distributed value chains, preclude the neat assignment of risks to a single quadrant \citep{r:91, r:92}. Instead, AI-related harms span low- and high-complexity dimensions and short- and long-term temporalities, producing a multidimensional spectrum of exposure that challenges conventional notions of risk allocation and system responsiveness, as well as any targeted intervention aimed at a specific risk category. This multidimensionality arises not only from AI’s technical capabilities but also from the architecture of the global AI ecosystem itself, which intertwines developers, deployers, end users, and supporting infrastructure across multiple jurisdictions and sectors. It is precisely these technological natures and the structure of the AI ecosystem that make transboundary AI risks likely to occur with greater frequency and severity in the future.

The transnational character of AI-related harm arises not merely from the cross-border movement of data or services \citep{r:45, r:46}, but from the distinctive architectural role AI systems, particularly foundation models, play within a globally distributed techno-ecosystem. Unlike previous digital technologies, such as cloud and computing infrastructures, whose transnationality is primarily organized around identifiable service providers, contractual relationships, and territorially anchored data centers, AI foundation models function as epistemic and functional substrates for a broad and constantly expanding array of downstream systems \citep{r:47, r:79, r:92}. Released under open-source or commercial licenses, these models are integrated, fine-tuned, and redeployed by heterogeneous actors across sectors such as healthcare, finance, logistics, public administration, and military operations \citep{r:47}. As a result, errors, biases, or emergent behaviors originating at the base model level propagate horizontally across applications and vertically through successive layers of adaptation, generating harms that are difficult to localize, either technically or legally. This mode of diffusion fundamentally differs from earlier forms of technological transnationality, in which cross-border effects were typically mediated by centralized intermediaries or clearly demarcated supply chains.

Globalized AI supply chains further intensify this form of boundary erosion. Training datasets, computational resources, model development, deployment, and post-deployment adaptation are routinely distributed across multiple jurisdictions, often without stable alignment between the location of development, use, and harm \citep{r:48, r:49, r:50}. Failures or misuses at any single node, whether in data curation, model training, fine-tuning, or contextual deployment, can cascade across jurisdictions, regardless of the originating actor’s size, intent, or regulatory environment. While cloud technologies similarly rely on distributed infrastructures, their risks tend to be infrastructure- and service-based, thereby channeling responsibility to identifiable providers. By contrast, AI’s general-purpose and modular nature enables a single foundational system to generate heterogeneous risks across multiple domains simultaneously, producing interconnected and compounding harms that are both individualized and systemic \citep{r:51}. This structural dispersion renders AI-related harm not only transboundary in effect but transboundary in origin, attribution, and governance, thereby exceeding the conceptual and institutional capacities of liability regimes designed for earlier generations of cross-border digital technologies.

Temporal dynamics and system adaptability also significantly expand the scope and heterogeneity of AI-related harms. Because AI systems continuously evolve through retraining, user interaction, and environmental feedback, they can exhibit emergent behaviors that transform low-probability risks initially into high-probability events \citep{r:12, r:52}. In long-lived deployments, particularly within critical infrastructure, public services, or essential economic systems, harms may accumulate hidden, gradually, producing slow-onset and difficult-to-detect effects as minor errors, biases, or inefficiencies compound over time \citep{r:53}. Simultaneously, these same systems can generate abrupt, high-impact failures because minor perturbations or unexpected inputs can propagate through complex, tightly coupled networks, triggering cascading effects or systemic instabilities. This coexistence of gradual accumulation and sudden disruption, combined with adaptive evolution, ensures that AI risk is inherently non-static: both the frequency and severity of adverse outcomes are likely to increase as AI systems proliferate across sectors, jurisdictions, and borders.

These technological characteristics interact with structural asymmetries in national capacities for AI development, oversight, and regulatory enforcement, thereby exacerbating global vulnerabilities, especially in downstream adopter states. Governments vary widely in their ability to audit, monitor, and govern complex AI systems, resulting in uneven exposure to risk \citep{r:54, r:55}. Jurisdictions with weaker regulatory frameworks or limited technical capacity may become inadvertent sources of harm, generating failures, misuse, or malicious exploitation that propagate internationally. Such spillovers are amplified when harmful data, model outputs, or behavioral patterns, whether produced intentionally or unintentionally, are incorporated into subsequent training cycles, embedding risks across multiple systems and jurisdictions. Consequently, even states with comparatively robust governance structures remain indirectly exposed due to the global entanglement of AI systems with cross-border data flows, shared infrastructures, and transnational supply chains.

Transnational threats are further intensified by the growing susceptibility of AI systems to cross-border attacks, including cyber sabotage, coordinated disinformation campaigns, and terrorist exploitation of AI-enabled infrastructures \citep{r:56, r:57}. As AI becomes embedded in logistics networks, surveillance systems, energy grids, and military or security decision-making, it increasingly presents high-value targets for non-state actors seeking to generate disproportionate disruption. Technological convergence accelerates these risks. The integration of AI with robotics, automation, cyber-physical systems, and high-performance computing produces tightly coupled hybrid infrastructures in which localized failures can cascade rapidly across sectors and regions. Errors in autonomous logistics, algorithmic financial trading, or critical infrastructure management may trigger systemic global effects. At the same time, AI-assisted decision-making in military or energy contexts can cause severe harm within seconds.

That is to say, these interdependencies ensure that both minor and catastrophic harms are not exceptional anomalies but recurring features of the evolving global AI ecosystem. Their scale, speed, and unpredictability are likely to intensify over time, as risks arise from the interplay among technical capabilities, deployment practices, governance asymmetries, and global interconnectedness. In this sense, transnational AI risks are structurally embedded within the system itself, rather than merely contingent on isolated failures or individual acts of misuse. As AI becomes increasingly integral to social, economic, and techno-systems, these cross-border risks are likely to grow both in frequency and impact, underscoring the need for proportionate attention and coordinated policy responses at the national and international levels.

\section{Governing Risk Beyond Borders: Comparative Approaches to Transnational AI Insurance}
While techno-uncertainties, particularly those arising from opaque AI systems, complicate efforts to anticipate and manage potential outcomes, it is precisely this uncertainty that has historically driven the evolution of risk mitigation and insurance mechanisms in human societies. In this paper, “insurance” is understood broadly to encompass a diverse array of practices and institutional arrangements for managing risk and distributing the burdens of recovery in the aftermath of accidents, disasters, or systemic failures. This conception extends well beyond conventional commercial or social insurance (both private and public, voluntary and mandatory) to embrace government-led or policy-driven mechanisms, such as taxation schemes, disaster relief programs, and statutory compensation frameworks, whose “insurance” function may be implicit rather than explicitly contractual. Accordingly, legal and regulatory frameworks operate as mechanisms of insurance: product liability regimes, tort systems, and broader compensatory justice structures establish ex ante rules, allocate responsibility, and create predictable pathways for redress, thereby embedding risk management directly into the architecture of society.

In this section, we examine historical cross-border insurance mechanisms, spanning both market-based and public-sector instruments, to understand how they have managed, mitigated, and adapted to diverse risk environments. We focus on the interaction between public and private institutions, highlighting how public mechanisms can create enabling conditions for private insurance and address risks that private markets cannot cover, while also noting the characteristics and limitations of each approach. 

\subsection{From Pandemics to Financial Crises: Rapid-Response Global Risk Governance}
From a political-economic perspective, governments have long had strong incentives to promote the development and cross-border integration of private insurance markets, leveraging the quasi-regulation capacity of private insurance firms. By facilitating the transfer and pooling of risk, regulating insurers, and managing systemic exposures that would otherwise fall directly on the state, governments stabilize insurance systems while mitigating potential fiscal and social liabilities \citep{r:58, r:59}. Such support has historically taken the form of domestic market incentives or the liberalization of entry for international insurers \citep{r:2}. In most sectors, once private actors are sufficiently incentivized, the state typically retreats from full liability, functioning instead as an “insurer of last resort” or, in some instances, abstaining from coverage entirely \citep{r:2, r:60}. This dynamic has rendered the growth of private insurance markets almost inevitable. Despite occasional failures, contemporary insurance and reinsurance markets generally operate with substantial depth and scale, even as their maturity, accessibility, and integration remain uneven across regions and nations.

In an increasingly uncertain and risk-prone world, the demand for security continues to expand the scope of the insurance industry. The emergence and proliferation of AI, with its novel and complex risks, provide particularly fertile ground for private insurance. In recent years, leading insurers and reinsurers, such as Lloyd’s \citep{r:22}, have begun developing products that cover AI-related exposures, with particular emphasis on cutting-edge technologies such as generative AI, reflecting growing market attention. The fundamental logic of insurance and reinsurance, combined with risk-transfer mechanisms such as catastrophe bonds and other Insurance-Linked Securities (ILS), allows risks to be migrated to capital markets, thereby expanding insurers’ capacity to underwrite additional exposures \citep{r:2}. These mechanisms suggest that a robust, transparent, and well-functioning private insurance sector can foster not only the adoption of novel technologies like AI but also the development of effective risk management and governance frameworks, with limited state intervention, serving either as a supportive backstop or as a last-resort guarantor.

However, a fundamental limitation arises from the nature of private insurance and reinsurance markets. These markets tend to operate most effectively in domains where risks are relatively foreseeable, liability structures are not excessively complex, potential harms are bounded in scope and duration, and the surrounding private insurance ecosystem is sufficiently mature to price and absorb such exposures \citep{r:2}. In the case of cross-border AI-related harms, this creates a distinct international vulnerability, reflecting differences in the maturity and readiness of domestic insurance markets and variations in legal frameworks across jurisdictions. Reliance solely on private markets fails to address the structural challenges posed by emerging AI technologies. First, it does not reconcile the difficulties AI introduces for fault-based liability systems, including the substantial evidentiary burdens placed on plaintiffs. Second, it does not overcome disparities in access to compensation mechanisms for third-party victims across countries. These gaps arise from differences in state capacity and the uneven development of national insurance markets, resulting in structural vulnerabilities for global risk management. In some cases, national private insurance markets may be underdeveloped, or governments may lack the capacity to support the private sector in covering inherently uninsurable risks, leaving victims with significant exposures unaddressed.

In such contexts, international and no-fault mechanisms provide a compelling model for addressing risks that are individual, accidental, and short-term, where private markets alone may be insufficient. Vaccine-related injuries illustrate this approach. During the COVID-19 pandemic, rapid vaccine development and deployment were essential to controlling the global health crisis, yet the urgency of mass immunization introduced significant uncertainties \citep{r:61}. New vaccines, including mRNA-based and other novel platforms, were authorized under accelerated timelines; although rigorously tested, they may not have captured all potential adverse effects. In this environment, traditional fault-based liability systems posed substantial challenges: plaintiffs faced high evidentiary burdens, lengthy litigation, and the risk of non-compensation, particularly in countries with weak legal infrastructures. At the same time, vaccine manufacturers faced the prospect of costly transnational liability claims that could slow production or limit global supply.

To mitigate these challenges, international no-fault programs, such as the Vaccine Injury Compensation Schemes implemented by Covax, WHO, and UNICEF, were established \citep{r:61, r:62}. These programs enabled rapid, predictable compensation by decoupling liability from fault and reducing procedural barriers. Crucially, they provided coverage even in conflict zones, areas of weak governance, or states lacking robust legal or insurance frameworks. By doing so, they addressed both equity, ensuring access for vulnerable populations, and efficiency, allowing manufacturers and public health agencies to deploy vaccines rapidly. In practical terms, such mechanisms allowed producers to prioritize speed of deployment over exhaustive local liability assessments while maintaining public trust, guaranteeing ex post compensation, and ensuring that potential harms did not impede global rollout \citep{r:62}.

Other examples demonstrate how urgent, coordinated action can similarly compel governments and international institutions to respond rapidly to complex, high-stakes risks. For high-complexity, rapidly unfolding hazards, including global financial crashes, timely interventions often require substantial state involvement, encompassing both voluntary international cooperation and multilateral institutional action. Historical precedents are instructive: the International Monetary Fund, empowered to stabilize exchange rates, coordinate reserve policies, and facilitate collective fiscal and monetary responses, exemplifies how states can jointly manage high-stakes, time-sensitive risks \citep{r:41}. These frameworks, both VICPs and IMF, function as risk-consolidation mechanisms. They centralize claims and remediation within a single institutional channel, substantially reducing, or in some cases entirely eliminating, the need for claimants to establish fault or engage in adversarial complaint-based procedures, focusing instead on the management and resolution of risk itself \citep{r:2, r:62}. By doing so, these mechanisms ensure that harms arising across heterogeneous legal systems are addressed in a timely, predictable, and equitable manner. In effect, they create institutional environments in which innovation and rapid deployment, whether in public health or financial stabilization, can proceed without leaving affected parties without recourse.

Yet, if such mechanisms are transposed to the AI domain without careful adaptation, significant normative and political–economic objections arise, particularly regarding the legitimacy of state-backed or intergovernmental no-fault compensation schemes. Importantly, the central point of contention does not concern the no-fault logic itself. On the contrary, a substantial body of scholarship endorses no-fault compensation as a suitable response to harms generated by emerging and opaque technologies such as AI, where causation is difficult to establish, responsibility is diffused, and ex post fault-based liability may be poorly aligned with the nature of risk \citep{r:11, r:12, r:13, r:63}.

Instead, the more contested issue lies in the role of the state or international organizations, whether national governments or IMF-analogous institutions, in underwriting such schemes. Unlike public health emergencies or global financial instability, where failures of collective action can generate immediate, severe, and system-wide harm, private actors predominantly drive AI development with significant commercial interests, and its outputs do not clearly constitute a public good in the same sense. In this context, the justification for public or intergovernmental backstopping is considerably less self-evident. State-supported compensation mechanisms risk socializing losses while leaving profits privatized, raising concerns about moral hazard and distributive inequity, or even leading to a "race to the bottom" in AI regulation.

Moreover, delayed or non-intervention by governments in AI-related incidents may not produce consequences as catastrophic or time-sensitive as those observed in pandemics or financial crises, further weakening the case for unconditional public coverage. Accordingly, while no-fault compensation may be normatively defensible for managing AI-related harms, far fewer scholars advocate for schemes that are fully underwritten by the state, except in circumstances where governments are deeply involved in, or directly benefit from, the risk-generating activities, such as through regulatory sandboxes \citep{r:63}, public–private development partnerships, or state-led deployment of AI systems.

This raises a critical design question: if space is preserved for private insurance markets to address cross-border AI-related risks, under what conditions should the state intervene as a backstop, and how should such intervention be structured to mitigate moral hazard while ensuring adequate compensation for victims? One potential model for consideration is the multi-layered insurance framework used in civil nuclear energy, in which private insurers provide initial coverage, national funds pool residual risk, and the inter-state collaboration acts as a final guarantor. 

\subsection{Governing Slow-Burn Catastrophes: Nuclear Power Accidents and Long-Horizon Liability}
In the aftermath of World War II, alongside the Marshall Plan’s strategy for postwar European reconstruction, U.S. energy policy increasingly turned to nuclear power as a potential engine of national and allied recovery, especially with the 1953 “Atoms for Peace” initiative of President Dwight D. Eisenhower \citep{r:64, r:65}. Unlike traditional energy sources, nuclear power promised both high profit margins and long-term strategic advantages, yet it carried low-probability, high-consequence risks capable of causing catastrophic, transnational, and long-lasting damage \citep{r:66, r:67, r:68, r:81}. These concerns were not merely hypothetical: subsequent nuclear disasters, including Three Mile Island, Chernobyl, and Fukushima Daiichi, vividly demonstrated the magnitude and complexity of potential hazards, highlighting the profound challenges inherent in managing nuclear risk \citep{r:69}.

Consequently, U.S. private enterprises were, for an extended period, markedly reluctant to invest in the nuclear sector, a hesitation rooted in a complex, multi-layered risk environment \citep{r:2}. The federal government was unwilling to assume unlimited liability, particularly while seeking to justify private-sector profits and to cover risks with public funds, and private insurers were similarly cautious, as potential damages were inherently unpredictable, could cross national borders, and might exceed the financial capacity of individual companies. The risk-sharing mechanisms available at the time were insufficient to reassure private actors that losses, including third-party claims exceeding corporate resources, would be effectively contained. In this context, private firms naturally refrained from entering the nuclear sector, making stagnation in the development of the U.S. nuclear industry, in many respects, almost inevitable, especially compared to the U.K. or the Soviet Union \citep{r:2}.

U.S. policymakers devised multi-tiered solutions, most prominently through the 1957 Price-Anderson Act. This legislation imposed strict liability on reactor operators while establishing a risk-sharing framework that combined private and public insurance, with the federal government serving as a backstop \citep{r:2, r:70, r:71}. At the first level, reactor operators were required to obtain private insurance coverage, capped at 50-65 million USD \citep{r:70}. Beyond that amount, another layer involved federal guarantees provided by the Atomic Energy Commission (AEC), which acted as a final guarantor of last resort, ensuring that even damages exceeding the combined private coverage could be addressed. The Act also encouraged the establishment of national risk-sharing funds administered by the American Nuclear Insurers (ANI), which coordinated coverage among both stock insurers and mutual insurers \citep{r:2}. This structure effectively distributed the financial risk of nuclear accidents across multiple private entities, ensuring that no single insurer would be overexposed while providing a predictable framework for compensation.

A comparable multi-tiered strategy has been reflected later in international nuclear liability regimes across the Atlantic, in the EU, particularly under the 1960 OECD's Paris (supported by Brussels) and 1963 UN's Vienna Conventions. The Paris Convention, for instance, establishes coverage ceilings that include private insurance (at least €700 million), national insurance (at least €500 million), and interstate-backed insurance (at least €300 million) \citep{r:2, r:72, r:73}. While the Vienna Convention does not mandate uniform ceilings, allowing member states to set their own limits, it similarly relies on layered public-private risk-sharing to ensure that nuclear damages are adequately addressed across both national and transnational contexts \citep{r:72, r:73}.

These international insurance mechanisms share several critical features. They impose strict channelling liability on reactor operators without requiring proof of fault and centralize claims through the operators themselves, thereby creating an exclusive channel for compensation \citep{r:74, r:75, r:76}. This design ensures that, as with the mechanisms of the VICPs or the IMF’s post-crisis management frameworks, there is a single, coordinated point of access for victims, thereby minimizing inconsistent compensation and conflicting interests and concentrating claims within the jurisdiction of the country where the incident occurred. These mechanisms also impose temporal limits on liability and compensation claims \citep{r:76}, as well as mandatory financial coverage, recognizing that private insurers’ capacity is inherently capped. In principle, these frameworks imply that private insurance alone cannot fully address the potential scale of nuclear liabilities. As a result, international markets must be organized through common funds or coordinated mutual arrangements among national insurance pools. This layered, state-backed architecture is grounded in empirical lessons from contemporary nuclear disasters, which demonstrate that if claims exceed private capacity, only governments, or coordinated intergovernmental structures, can reliably ensure full compensation and maintain systemic stability.

However, these mechanisms are not without limitations, particularly when applied to AI liability insurance. In principle, the international frameworks developed for nuclear security could provide a conceptual model for AI, a topic that scholars are increasingly examining \citep{r:39, r:93, r:94}. Yet a closer look at the tensions that arose during the drafting, implementation, and enforcement of these international conventions reveals the potential difficulties of creating a comparable system for AI. Disputes may emerge over how to define an “AI incident” eligible for insurance coverage, as well as how to calibrate liability and coverage levels for incidents of varying severity.

Mandatory private and national coverages, analogous to the Paris Convention’s thresholds of at least €700 million in private insurance and €500 million in state-backed coverage, could, in theory, exclude countries with limited capital, regions with weak or absent governance, or economies with underdeveloped private insurance markets. By contrast, the Vienna Convention, which permits states to set their own national ceilings, illustrates the challenges posed by uneven capacity: disparities in resources and regulatory frameworks can generate tensions over the allocation of responsibility. Until the late 1980s, the Paris and Vienna regimes operated independently \citep{r:21, r:77}, meaning that a nuclear accident in a Paris Convention country did not automatically entail liability for related damages in a Vienna Convention state. Controversies emerged over whether more cautious countries should bear responsibility for mismanagement elsewhere, or whether a state with a minor or nonexistent nuclear sector should be held liable for facilities that are larger and riskier, a question further complicated by the existence of atomic operators outside both regimes, such as China \citep{r:2}. A similar debate may emerge in AI, where the scale of development, deployment, and infrastructure varies dramatically across nations. 

Strict liability, channeling, and exclusive liability, which have been central features of international nuclear liability regimes, pose another significant challenge when considered in the context of cross-border AI governance. Civil nuclear liability conventions uniformly assign responsibility for damages to the reactor operator, including accidents that occur during the transport of radioactive materials. Debates over the scope and limits of operator liability have historically generated prolonged tensions, particularly in contexts where no convention exempts operators from responsibility in the event of deliberate attacks or terrorism, effectively anchoring the duty of protection with the facility itself. The Paris Convention similarly rejects force majeure exemptions, as exemplified in its 2004 clarification regarding natural disasters, armed conflict, hostilities, or civil war \citep{r:2, r:78}. In principle, these rules rest on the assumption that nuclear operators are best positioned both to internalize risk and to implement comprehensive, adequate safeguards to prevent accidents, thereby creating strong incentives for proactive risk management.

Yet the drafting process of the Paris Convention, notably contested by Germany and Austria, highlights an opposing but equally principled concern: channeling all risk to the operator not only exceeds the facility’s inherent capacity but also prioritizes efficiency in accident prevention and compensation over notions of individual or collective justice \citep{r:2}. The no-fault structure, while effective in concentrating risk management, implicitly precludes holding negligent parties accountable, raising questions about the balance between systemic safety and moral responsibility \citep{r:2}. Analogous tensions are likely to emerge in AI if strict liability and responsibility channeling are applied. Key debates would include whether strict liability is an appropriate mechanism when risks are inherently unpredictable and beyond the preventive capacity of AI developers or deployers, and whether assigning comprehensive responsibility to such entities would generate counterproductive chilling effects that slow the development of an industry of strategic importance. Equally critical is the question of where responsibility should be channeled. In nuclear energy, liability is concentrated in extensive, capital-intensive operation facilities with significant capacity to absorb risk. In AI, by contrast, downstream technology companies, including small or medium-sized enterprises or startups \citep{r:97, r:98, r:99}, particularly in middle- and lower-income countries, may lack the capacity to internalize systemic risk \citep{r:95, r:96}. Determining the appropriate locus for channeling liability within a fragmented AI ecosystem, where responsibilities are diffuse and contributions difficult to trace, represents a profound governance challenge, arguably bordering on the practically infeasible.

Extended temporal horizons for the emergence of damages introduce additional and profound governance challenges. In nuclear energy, latent and initially imperceptible losses, such as those caused by long-term radioactive exposure, have historically complicated claims and liability allocation. Under the Paris Convention, for example, property damage claims are generally limited to ten years, though the 2004 Protocols, implemented by many signatory states, extend personal injury claims to thirty years to account for delayed health effects \citep{r:78}. In the context of AI, analogous but potentially far more complex harms may arise, encompassing environmental impacts, cultural erosion, algorithmic bias, and other socially and politically contested outcomes. These damages are often diffuse, cumulative, and difficult to attribute to a single actor, which amplifies both uncertainty and contestation.

As in environmental governance, slow-onset or gradual disasters may be obscured by short-term economic and social priorities, intensifying public debate and political disagreement. This dynamic is particularly evident in transnational contexts, such as international climate policy, where disagreements over responsibility, enforcement, and resource allocation create significant obstacles to cooperation. Similarly, in AI governance, divergent national capacities, economic interests, and political priorities, compounded by the involvement of non-state actors, could impede the establishment of coordinated, effective mechanisms for monitoring, liability, and remediation. 

Placed within the context of U.S.–China technological competition, these challenges are further magnified: disparities in AI development capabilities, strategic incentives to maintain national advantage, and asymmetries in regulatory transparency may foster strategic mistrust, making multilateral agreements on liability or insurance difficult to negotiate. Moreover, the prioritization of national security and economic dominance could incentivize countries to externalize risk, leaving transboundary harms unaddressed, while differences in legal systems and enforcement capacity could exacerbate conflicts over compensation and accountability. These temporal and cross-jurisdictional dimensions suggest that designing robust liability and insurance frameworks for AI will require careful attention not only to technical risks but also to political, economic, and societal factors that shape the feasibility of international cooperation.

\section{Conclusion}
This paper has examined the complexity of transboundary AI risks and their potential to escalate in scale, frequency, and societal impact. Historical precedents, including public–private insurance arrangements, state-backed private coverage, no-fault vaccine injury compensation schemes, IMF crisis interventions, and layered international liability regimes in civil nuclear energy, offer instructive, if imperfect, analogies for the design of AI risk governance mechanisms.

Several features of these models are potentially relevant to AI. Strict liability regimes can centralize responsibility and strengthen preventive incentives, providing an alternative to fault-based approaches ill-suited to opaque and unpredictable AI behavior, though they raise concerns about fairness and innovation deterrence. Collective compensation funds may help address diffuse, delayed, or attribution-resistant harms, but pose challenges of adequate capitalization and the risk of public subsidization of private gains. Responsibility-channeling mechanisms reduce coordination costs and litigation complexity by designating a single accountable actor, yet may overburden entities with insufficient capacity. Finally, shared international risk-pooling arrangements could enhance systemic resilience, but their feasibility is constrained by fragmentation and geopolitical competition in the AI ecosystem. Taken together, these mechanisms involve persistent trade-offs among efficiency, fairness, and political feasibility, underscoring that no single model offers a comprehensive or uncontested solution.

Given the descriptive and analytical orientation of this research, the discussion remains necessarily conceptual rather than fully operational. This limitation reflects the intrinsic uncertainty of AI technologies, their rapid evolution, and the contested nature of transboundary responsibility, all of which are compounded by global asymmetries in AI capabilities, regulatory capacity, and strategic incentives, particularly among major powers. These conditions underscore the political as well as technical barriers to effective governance. Yet, by systematically drawing on lessons from analogous high-risk, cross-border domains, this research provides a foundation for forward-looking deliberation. It suggests that effective AI governance and liability regimes will likely require multilayered frameworks combining strict liability, pooled compensation mechanisms, targeted responsibility channeling, and international risk-sharing arrangements. While such frameworks will remain imperfect and politically contested, early and deliberate design can mitigate catastrophic risks, incentivize responsible development and deployment, and enable iterative policy adaptation in a domain where the stakes are inherently global.

\bibliographystyle{plainnat}
\bibliography{Tran}

\end{document}